\documentclass{appolb}
\usepackage{graphicx}

\begin{document}
\title{Exotic atoms and exotic nuclei%
}
\author{S. Hirenzaki, H. Nagahiro
\address{Department of Physics, Nara Women's University, Nara 630-8506, Japan}
\\
{  
}
\vspace{5mm}
N. Ikeno
\address{Tohoku University, 6-3 Aoba, Aramaki, Aoba, Sendai, 980-8578 Miyagi, Japan \\
YITP, Kyoto University, Kyoto 606-8502, Japan
}
\\
{  
}
\vspace{5mm}
J. Yamagata-Sekihara
\address{RCNP, Osaka University, Ibaraki, Osaka 567-0047, Japan}
}
\maketitle
\begin{abstract}
We briefly review the study of the exotic atoms and exotic nuclei, and report recent research activities of $\eta$-mesic nucleus and kaonic atoms in this article.  
\end{abstract}
  
\section{Introduction}
Exotic atoms and exotic nuclei are one of the new kinds of the sub-atomic matter which include exotic particles such as mesons, excited baryons and so on.  The study of these systems is one of the important subjects in the contemporary hadron-nuclear physics.  The detailed study of the structure of such systems can provide the precise information on the hadron-nucleus interaction and can be considered as the natural extension of the research filed of the nuclear physics to new frontiers.  

In addition, we also believe that we can deduce the information on the aspects of the symmetry of the strong interaction at finite density from the study of the exotic atoms and exotic nuclei \cite{Yamazaki12}.  In the standard scenario believed by the majority of hadron physicists, the chiral symmetry is spontaneously broken in the vacuum and is expected to be restored at high density and/or temperature circumstances \cite{HatsudaKunihiro}.  The exotic systems we consider here are expected to provide the finite density circumstance to mesons and provide the chance to obtain the property change of mesons which could have close connection to the symmetry aspects, partial restoration of chiral symmetry at finite density.

In these contexts, various exotic systems have been investigated.  For example,  the change of the chiral condensate in nucleus is concluded in the study of the deeply bound pionic atoms \cite{Suzuki04}. The  recent theoretical and experimental developments can be found in Refs.  \cite{Ikeno11,Ikeno11-2,Ikeno13, Itahashi2014,ItahashiPC2}.
And new  attempts to deduce information on $U_{\rm A} (1)$ anomaly effects at finite density from the study of the $\eta'$(958) mesic nuclei have started, too \cite{nagahiro05, nagahiro06, jido12, nagahiro12, itahashi12, nagahiro13}.  Within the various systems, we briefly report recent topics on $\eta$ mesic nucleus and kaonic atoms in this article.  

\section{$\eta$-mesic nuclei}

Bound states of the $\eta$ meson in nuclei were predicted by Haider and Liu in 1980's \cite{Haider86},  and after that many works were devoted to the studies of the structure and formation of these states.  
Especially, the study of the $\eta$-mesic nucleus based on the chiral symmetry has been developed significantly \cite{jido02,Inoue02}. Due to the strong coupling of $\eta N $ to $N^*(1535)$ resonance, the $\eta$ mesic nuclei are expected to provide the unique chance to study the aspects of chiral symmetry of baryon sector since $N^*(1535)$ is the lightest non-strange baryon with opposite parity to nucleon and is a candidate of the chiral partner of nucleon \cite{jido02}.  
 
The first experiments for the formation of the $\eta$ mesic nucleus was performed in 1988 by the $(\pi,p)$ reaction \cite{Chrien} and the results were turned to be negative.  The experiment was performed at the finite proton angle in the laboratory frame to observe narrow peak structure predicted at that time.   The $(\pi,p)$ spectra are calculated recently \cite{nagahiro2009} with the same condition with the experiment \cite{Chrien} and the results are shown in Fig. 1. 
Two theoretical models have been used to obtain the $\eta$-nucleus interaction, which are based on the much different picture of $N^*(1535)$. 
As we can see from the figure, 
the narrow peak structure in the spectrum does not appear in these theoretical predictions. And we can also find that it is hard to distinguish two theoretical models from the existing data.  

So far, there have been no experimental data which can prove the existence of the $\eta$-nucleus bound states.  Thus, we think that we need further experimental and theoretical studies of the $\eta$-nucleus systems.  Especially, the formation of the $\eta$ mesic nucleus in the light nucleus like He \cite{Skurzok:2011aa, Krzemien:2012dd, Adlarson:2013xg, Krzemien:2014qfa}
is interesting since the width of the system could be large as predicted by some theoretical results \cite{jido02,Inoue02}.

\begin{figure}[h]
\begin{center}
\includegraphics[scale=1]{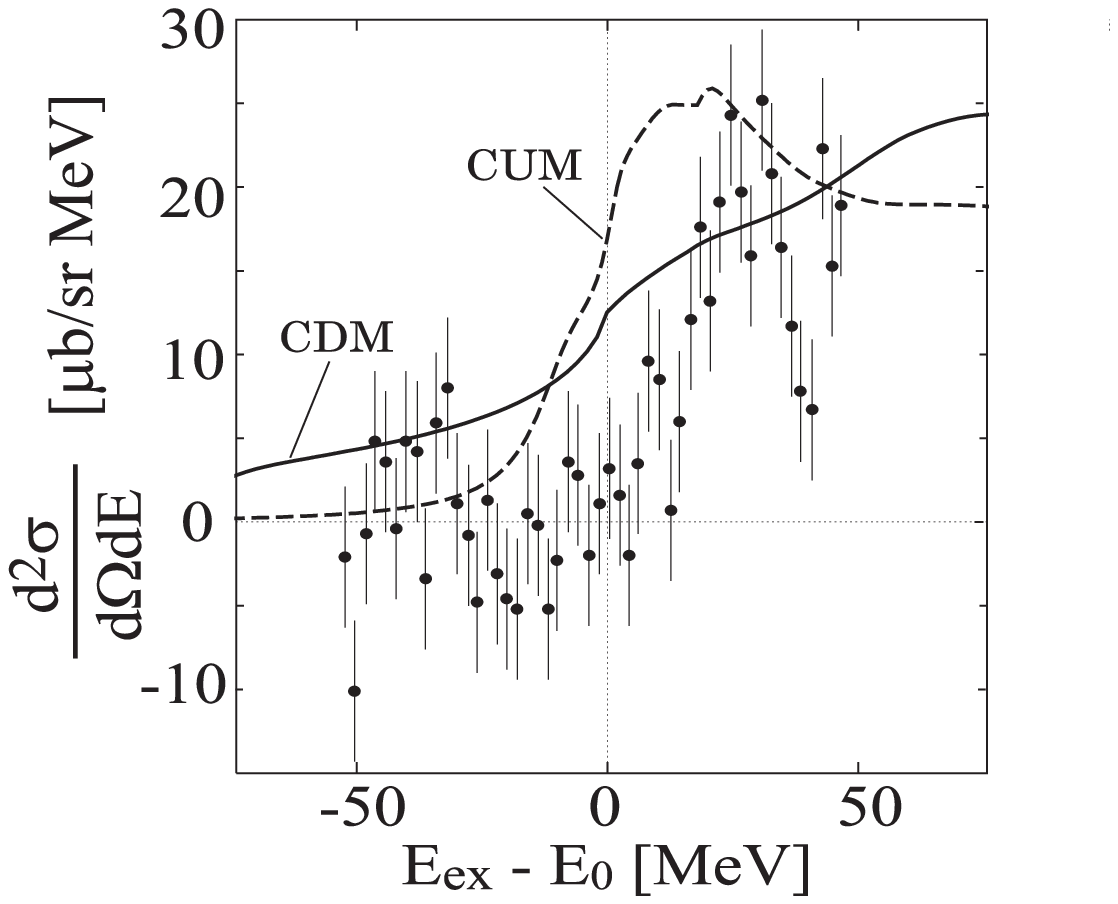}
\end{center}
\caption{Comparison of the calculated spectra \cite{nagahiro2009} and the experimental data \cite{Chrien} on the carbon target case reported in Ref. \cite{nagahiro2009}. The background shown in Ref. \cite{Chrien} is subtracted. The solid line indicates the total spectrum with the chiral doublet model (CDM) and the dashed line is that of the chiral unitary model (CUM). See details in Ref. \cite{nagahiro2009}.  }
\label{Fig:F1}
\end{figure}

\section{Kaonic atoms}

Kaonic atoms have been believed to provide important information concerning the $K^-$-nucleon interaction in the nuclear medium and the $K^-$-nucleus interaction.  This information is very important in understanding kaon properties at finite density, which has close connection to the  structure and formation of the kaonic nuclei \cite{AY, dote04} and the constraints on kaon condensation in high-density matter.  

The $K^-$-nucleus interaction has been studied for a long time based on the lightly bound kaonic atom data obtained by the x-ray spectroscopy.  In Ref. \cite{BFG97}, a phenomenological study of the kaonic atoms was performed comprehensively and the potential shape and strength were determined by the $\chi^2$ fit to the data. On the other hand, there have been significant developments in the theoretical description of hadron properties in terms of the chiral Lagrangian \cite{KSW,ramos2000}.  Based on the theoretical framework, 
the kaon self-energy in the nuclear medium has been obtained theoretically, which can be tested with the kaonic atom data \cite{ramos2000}.   The theoretical potential is known to reproduce the experimental data reasonably well \cite{Hirenzaki00}.  

The one of the biggest puzzles in the study of the $K^-$-nucleus interaction is the significantly large discrepancy ( $\sim$ factor 3) between the potential strengths of the phenomenological potential \cite{BFG97} and the theoretical potential \cite{ramos2000}.  We show the level structure of the kaonic atoms in $^{207}$Tl reported in Ref. \cite{yamagata2007} in Fig. 2 as an example. As reported in Refs. \cite{yamagata2007,yamagata2005}, the results obtained by both potentials resemble each other though the potential strengths are much different as mentioned above.  This behavior of the level structure can be understood by considering the kaonic nuclear states expected to exist inside the nucleus.  The number of the nuclear state inside nucleus is different for these two potentials and, as the consequence, the atomic level structure can be the same for the potentials with significantly different depths.  

Recently, there have been the significant developments in the experimental accuracy for the observation of the kaonic x-ray and the extremely high precision data could be obtained in near future \cite{okada2014}.  The high resolution data could be used to disentangle the puzzle of the depth of the $K^-$-nucleus interaction and to obtain new information on kaonic nuclear states.  We evaluated theoretically the difference of the x-ray energy between the phenomenological and theoretical potentials for $3d \rightarrow 2p$ transition of kaonic helium to be about 0.3 eV \cite{yamagata14}.  The difference of the level widths due to the strong absorption of kaon is also evaluated to be about 1 eV between two potentials \cite{yamagata14}.  These tiny differences could be observed in the new experiments in near future \cite{okada2014}, and it will be an important step for the study of kaon nucleus interaction.

\begin{figure}[h]
\begin{center}
\includegraphics[scale=1.3]{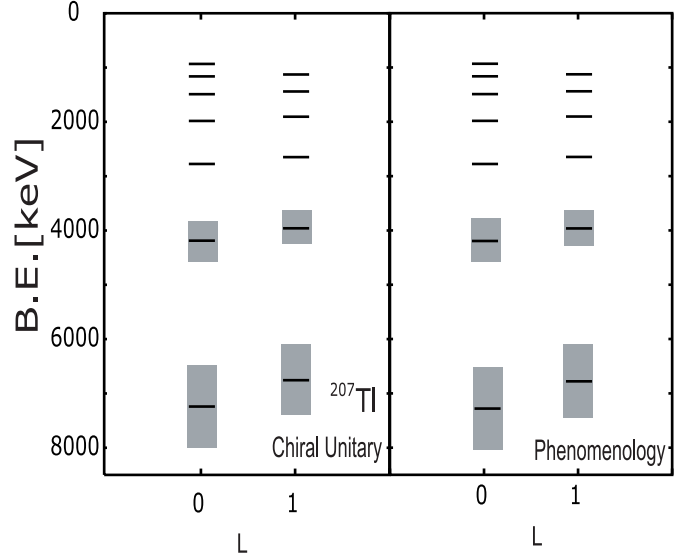}
\end{center}
\caption{Energy levels of kaonic atoms of $^{207}$Tl up to the principle quantum number $n$ = 7 for $s$ and $p$ states obtained with the theoretical optical potential of the chiral unitary model (left) and of the phenomenological fit (right) \cite{yamagata2007}. The hatched areas indicate the level widths for deeply bound states (B.E. $>$ 3 MeV).  }
\label{Fig:F2}
\end{figure}

\section{Summary}

We have briefly reviewed the recent research activities of $\eta$-mesic nucleus and kaonic atoms in this article.  We think the study of the exotic atoms and the exotic nuclei will provide new frontier of the hadron many body physics and also provide simultaneously the  finite density circumstance for hadrons which can be used for the study of the aspects of the strong interaction symmetry in medium.  In this sense, the study of these systems are expected to be fruitful and should be developed both theoretically and experimentally.


\end{document}